\documentclass[12pt]{iopart}

\usepackage{graphicx}
\usepackage{siunitx}
\usepackage{bm}
\usepackage{braket}
\usepackage{hyperref}

\begin{document}
\title{Expansion of a quantum gas in a shell trap}

\author{Yanliang Guo$^1$\footnote{Present address: Institut für Experimentalphysik und Zentrum für Quantenphysik, Universität Innsbruck, 6020 Innsbruck, Austria},
Emmanuel Mercado Gutierrez$^2$, David Rey$^1$, Thomas Badr$^1$, Aurélien Perrin$^1$, Laurent Longchambon$^1$,
Vanderlei Salvador Bagnato$^2$, Hélène Perrin$^1$ and Romain Dubessy$^1$
}

\address{$^1$ Université Sorbonne Paris Nord, Laboratoire de Physique des Lasers, CNRS UMR 7538, F‐93430, Villetaneuse, France}
\address{$^2$ Instituto de Física de S\~ao Carlos, Universidade de S\~ao Paulo, CP 369, S\~ao Carlos, S\~ao Paulo, 13560-970, Brazil}
\ead{romain.dubessy@univ-paris13.fr}

\begin{abstract}
We report the observation of the controlled expansion of a two-dimensional quantum gas confined onto a curved shell-shaped surface.
We start from the ellipsoidal geometry of a dressed quadrupole trap and introduce a novel gravity compensation mechanism enabling to explore the full ellipsoid.
The zero-point energy of the transverse confinement manifests itself by the spontaneous emergence of an annular shape in the atomic distribution.
The experimental results are compared with the solution of the three-dimensional Gross-Pitaevskii equation and with a two-dimensional semi-analytical model.
This work evidences how a hidden dimension can affect dramatically the embedded low-dimensional system by inducing a change of topology.
\end{abstract}
\submitto{\NJP}
\maketitle
%\tableofcontents
\section{Introduction}
When the motion of a dynamical system is constrained within a particular domain new effects may occur. In particular if one or more degrees of freedom are frozen the system can be described by an effective low dimensional theory~\cite{DaCosta1981}. For example the \emph{classical} rigid pendulum oscillates in a two-dimensional (2D) plane but is described by an effective one-dimensional (1D) equation.
In the \emph{quantum} world numerous examples exploit this possibility to obtain new effects, as for example the realization of 1D channels~\cite{Krinner2015}, mesoscopic quantum devices~\cite{Imry1998} or Hall effect in 2D electron gases~\cite{Hall1879}.
Any physical low dimensional system is still embedded in a higher dimensional space whose properties can affect the motion. For example the curvature of the constrained surface is expected to give rise to additional potential terms\cite{DaCosta1981,Kaplan1997,Sandin2017a}, while the inhomogeneity of the confining potential contributes through a slow variation of the zero point energy~\cite{Krinner2015,Sandin2017a,Schwartz2006a}.

Ultracold atom experiments offer a unique playground to probe lower dimensions~\cite{Gorlitz2001,QGLD2003}, with many impressive achievements, as for example the simulation of the 1D Lieb-Liniger model~\cite{Kinoshita2006b}, the observation of the 2D Berezinski Kosterlitz Thouless model~\cite{Hadzibabic2006,Fletcher2015} or the possibility to realize synthetic dimensions~\cite{Mancini2015,Chalopin2020}. They recently enabled the discovery of new dynamical effects in 2D~\cite{Saint-Jalm2019a,Shi2020,Olshanii2021}. For interacting systems, the effective dimension can change the nature of interactions~\cite{Olshanii1998,Olshanii2010} which in turn modifies the equation of state~\cite{Hung2011a,Merloti2013a}.
Recently, the original topology of a Bose-Einstein condensate (BEC) spread onto a closed spherical surface has motivated several theoretical studies~\cite{Bereta2019,Tononi2019,Tononi2020,Moller2020a,Bereta2021}, and an experiment aiming at this goal is currently installed in the International Space Station \cite{Lundblad2019}.

\begin{figure}[t]
\centering
\includegraphics[width=7cm]{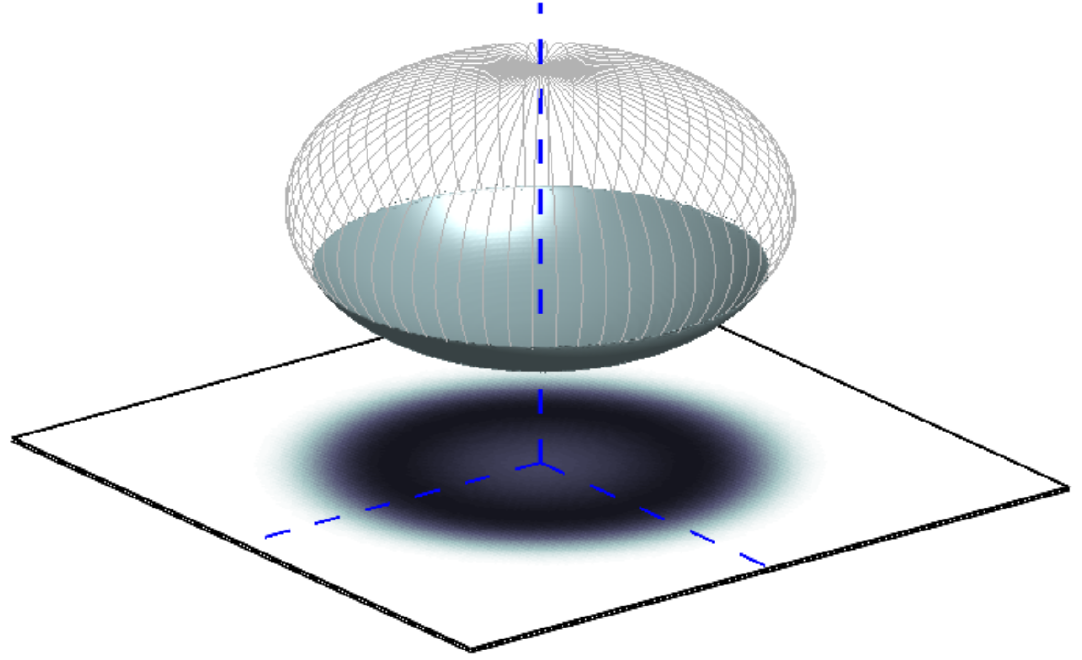}
\includegraphics[width=7cm]{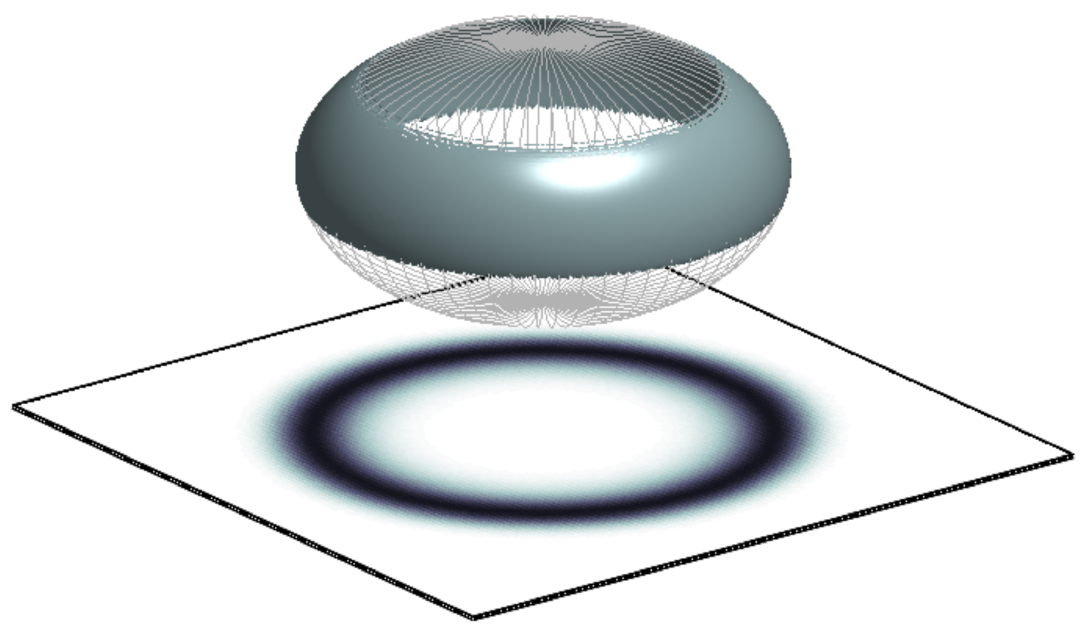}
\caption{\label{fig:3D}
Sketch of the experiment reported in this work: a quantum gas (light grey annulus) is constrained to move on a spheroidal surface (meshed surface). Left: thanks to gravity compensation the atoms explore a large fraction of the surface. Right: when gravity is over compensated the atoms accumulate at a given height because of the inhomogeneous transverse zero-point energy, see text for details. The shadow at the bottom shows the integrated density distribution of the gas, blurred by a $\SI{4}{\micro\metre}$ point spread function to reproduce the experimental imaging resolution. The field of view is $\SI{120}{\micro\metre}\times\SI{120}{\micro\metre}$.
}
\end{figure}

In this article, we report the direct observation of the effect of the inhomogeneous zero-point energy on a gas confined to an ellipsoid surface. 
A novel gravity compensation mechanism enables the  exploration of the full ellipsoid in the spirit of the space experiment of~\cite{Lundblad2019}. We demonstrate how the motion restricted to the surface is strongly affected by the transverse frozen degree of freedom, resulting in an annular shape as shown in \fref{fig:3D}. Our work illustrates how the inhomogeneity of the underlying three-dimensional potential can induce a change of topology in the effective 2D Hamiltonian and how this effect become predominant in a pseudo-microgravity environment. In contrast to most experiments where it is only a small correction to the external potential, here the quantization of the transverse motion is central to the realization of an annular gas.

To constrain the motion of the atoms to a surface, we make use of adiabatic potentials realized with radio-frequency (rf) dressed ultracold atoms.
They allow to access a variety of trapping geometries~\cite{Garraway2016a} from double wells~\cite{Schumm2005b,Hofferberth2006,Lesanovsky2006,Barker2020} to bubble traps~\cite{Colombe2004a} and even reach the two-dimensional regime~\cite{Merloti2013a}.
Thanks to the high degree of control on all parameters they are ideally suited to study superfluid dynamics~\cite{Merloti2013b,Dubessy2014,DeRossi2016,Guo2020}.
By time averaging~\cite{Lesanovsky2007a,Sherlock2011} or multiple dressing~\cite{Harte2018} even more configurations can be realized, as smooth ring-shaped waveguides~\cite{Navez2016,Pandey2019} or multiple wells~\cite{Harte2018}. 

\section{Gravity compensation in a shell shaped trap}

Our experimental setup is described in~\cite{Merloti2013a,Dubessy2012a}.
Briefly, $^{87}$Rb atoms in the $F=1$ ground state are placed in a rotationally invariant quadrupole magnetic field of main vertical axis $z$. 
The atoms are dressed by an rf field produced by three antennas with orthogonal axes, fed by a homemade direct digital synthesis device, allowing for a full control of the rf polarization and a fine tuning of its parameters.
The resulting potential, within the rotating wave approximation (RWA), can be derived following the approach introduced in~\cite{Zobay2001} (for a review see also \cite{Perrin2017}) and reads~\cite{Merloti2013a}:
\begin{equation}
V^{\rm RWA}_{\rm 3D}(\rho,\phi,z)
=\hbar\sqrt{(\alpha\ell-\omega)^2+\Omega(\rho,\phi,z)^2}+Mgz,
\label{eqn:potential}
\end{equation}
where $\ell^2=\rho^2+4z^2$, $(\rho,\phi,z)$ are the usual cylindrical coordinates, $\alpha$ is the quadrupole gradient in the horizontal plane in units of frequency, $\omega$ is the rf frequency, $\Omega(\rho,\phi,z)$ is the local atom-field coupling amplitude and the last term is the gravitational potential.
$\Omega(\rho,\phi,z)$ depends on the orientation of the rf polarization with respect to the local static magnetic field \cite{Perrin2017}. 
For the choice of a circular polarization of axis $z$, $\Omega(\rho,\phi,z)=\Omega_0/2\times(1-2z/\ell)$ and the potential is rotationally invariant.
Hereafter we will drop the explicit $\phi$ dependence in all quantities.
The locus of the energy minimum in equation~\eref{eqn:potential} belongs to a 2D ellipsoidal isomagnetic surface, defined by $\ell=r_0\equiv\omega/\alpha$, slightly deformed by the gravitational sag~\cite{Merloti2013a}.
We note that $\Omega(\rho,z)$ reaches its maximum value $\Omega_0$ at the bottom and vanishes at the top of this surface.

\begin{figure}[t]
\centering
\includegraphics[width=\textwidth]{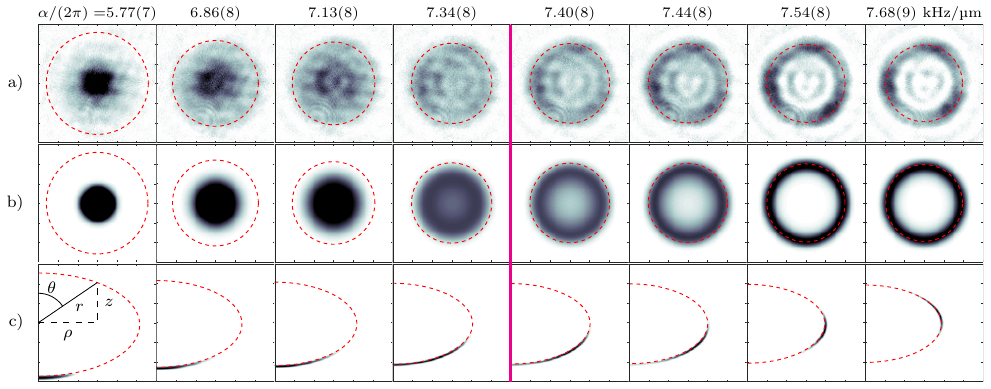}
\caption{\label{fig:ramp}
(color online)
In situ atomic density distribution for an ensemble of $N\simeq10^5$ atoms, evidencing the gravity compensation mechanism and the spontaneous change of topology, as the quadrupole gradient $\alpha$ increases. 
a) experimental measurement, b) and c) full GP numerical simulation, top and side views respectively. The pink vertical line corresponds to the observed threshold for gravity compensation, slightly lower than the naive expectation $\alpha_g/(2\pi)=\SI{7.54\pm0.04}{\kilo\hertz/\micro\metre}$, see text for details. For each picture of a) and b) the field of view is $\SI{120}{\micro\metre}\times\SI{120}{\micro\metre}$, the color scale spans $[0-35]\,\SI{}{\micro\metre^{-2}}$, and the dashed red circles indicate the ellipsoidal radius at equator $\rho\equiv r_0$. b) The simulated density profiles are convoluted with a Gaussian of $1/\sqrt{e}$-radius $\sigma=\SI{4}{\micro\metre}$ to reproduce the experimental imaging resolution.
For c) the field of view is $\SI{60}{\micro\metre}\times\SI{60}{\micro\metre}$ and the dashed red line is the adiabatic surface $r=r_s(\theta)$, see text for details.
}
\end{figure}

The confinement to this ellipsoid is rather strong. If we assume that the atoms are confined to the ground state of the motion transverse to the surface, we can derive an expression of the potential for the 2D motion along the ellipsoid.
If we neglect the deformation of the surface due to the gravitational sag, we obtain a simple expression for the effective 2D potential:
\begin{equation}
V^{\rm RWA}_{\rm 2D}(z)=\frac{\hbar\Omega_0}{2}+\left(Mg-\frac{\hbar\Omega_0}{r_0}\right)z+\frac{\hbar\omega_\perp(z)}{2},
\label{eqn:Vtoy}
\end{equation}
where the atoms move on the isomagnetic surface, and we have used $\rho^2=r_0^2-4z^2$, for $|z|\leq r_0/2$ in equation~\eref{eqn:potential}.
In this expression we have neglected the geometrical effect of the curvature on the potential~\cite{DaCosta1981}, resulting in a energy difference of order $\hbar^2/(Mr_0^2)$ between the poles and the equator, about $h\times\SI{0.1}{\hertz}$ with our parameters.
Equation~\eref{eqn:Vtoy} shows that the inhomogeneity in rf coupling amplitude results in a force acting against gravity.
Gravity can be compensated by an appropriate choice of the magnetic field gradient, fulfilling $\alpha_g= Mg\omega/(\hbar\Omega_0)$.
The last term in equation~\eref{eqn:Vtoy} involving the transverse confinement frequency $\omega_\perp(z)$ is a witness of the higher dimension, entering through the zero-point energy of this degree of freedom~\cite{Sandin2017a,Schwartz2006a}.
It scales as $\alpha(z)/\sqrt{\Omega(z)}$, where $\alpha(z)$ and $\Omega(z)$ are the local gradient and coupling respectively.
This quantum effect is responsible for the spontaneous change of topology when gravity is overcompensated, as shown in \fref{fig:ramp}: the zero-point energy contribution to the effective potential becomes dominant as the atoms are pushed towards the top and $\Omega(z)$ vanishes.

To demonstrate this effect we initially load the adiabatic potential with a moderate gradient $\alpha/(2\pi)=\SI{4.14\pm0.06}{\kilo\hertz/\micro\metre}$ and a circularly polarized rf dressing field of frequency $\omega/(2\pi)=\SI{300}{\kilo\hertz}$ and maximum coupling amplitude $\Omega_0/(2\pi)=\SI{85.0\pm0.5}{\kilo\hertz}$.
A radio frequency knife of frequency $\omega_{\rm kn}/(2\pi)=\SI{104}{\kilo \hertz}$, linearly polarized along the vertical axis, allows to control the trap depth while preserving the rotational symmetry.
We then increase the gradient within $\SI{300}{\milli\second}$ while keeping all the other parameters constants, and record an in-situ picture of the atomic density distribution using a standard absorption imaging scheme with the probe beam propagating along the $z$ axis~\cite{Merloti2013a}.

\Fref{fig:ramp} shows that for increasing values of the gradient $\alpha$, the atomic cloud expands progressively to fill the ellipsoidal surface and, when gravity is overcompensated, i.e., for $\alpha/(2\pi)\geq\SI{7.40\pm0.08}{\kilo\hertz/\micro\metre}$, takes a stable annular shape close to the equator. Interestingly, the compensation occurs for a gradient slightly lower than the naive expectation $\alpha_g=2\pi\times\SI{7.54\pm0.04}{\kilo\hertz/\micro\metre}$.
A correct modeling of the system, including beyond RWA correction to equation~\eref{eqn:potential} and an exact description of the frozen degree of freedom is necessary to obtain the quantitative agreement shown in \fref{fig:ramp} between theory and simulation, as detailed below.
We emphasize that the gradient $\alpha$ and the coupling $\Omega_0$ are calibrated with independent measurements, see~\ref{app:methods}, and that there is no free parameter in the simulations shown in \fref{fig:ramp}.

\section{Effective two-dimensional potential}

In order to refine the theoretical description of the ring formation, we first use a Floquet (Fl) expansion~\cite{Shirley1965,Hofferberth2007a} to include beyond RWA terms, see \ref{app:Floquet} for details. 
We find that even for our moderate coupling amplitude $\Omega_0/\omega=0.28$ it is necessary to include the first five manifolds, up to $\pm2$ photons, to reach convergence in the computation of the adiabatic potential $V^{\rm Fl}_{\rm 3D}(r,\theta)$. 
Here, $(r,\theta,\phi)$ are the spherical coordinates, see \fref{fig:ramp}c, and $V^{\rm Fl}_{\rm 3D}$ does not depend on $\phi$.
Using this more accurate potential, we compute numerically the mean-field atomic wave function  with the Gross-Pitaevskii (GP) equation.
The GP equation is propagated in imaginary time on a discrete grid to obtain the three-dimensional ground state~\cite{Antoine2017}.
We exploit the rotational invariance to speed up the computation and use a map to ellipsoidal coordinates to 
achieve good accuracy at every point of the surface~\footnote{The details will be published elsewhere.}.

Finally, we develop an improved semi-classical two-dimensional description of the potential restricted to a surface, improving the accuracy of equation~\eref{eqn:Vtoy}.
For each angle $\theta$ we compute the potential $V^{\rm Fl}_{\rm 3D}(r,\theta)$ and find its minimum as a function of $r$, thus defining the constrained surface $r=r_s(\theta)$.
For each point of this surface we also compute the local Hessian matrix and obtain the transverse confinement frequency $\omega_\perp(\theta)$ from its largest eigenvalue.
The improved semi-classical 2D potential reads:
\begin{equation}
V^{\rm Fl}_{\rm 2D}(\theta)=V^{\rm Fl}_{\rm 3D}(r_s(\theta),\theta)+\frac{\hbar\omega_\perp(\theta)}{2}.
\label{eqn:Vimp}
\end{equation}

\begin{figure}[t]
\centering
\includegraphics[width=8.6cm]{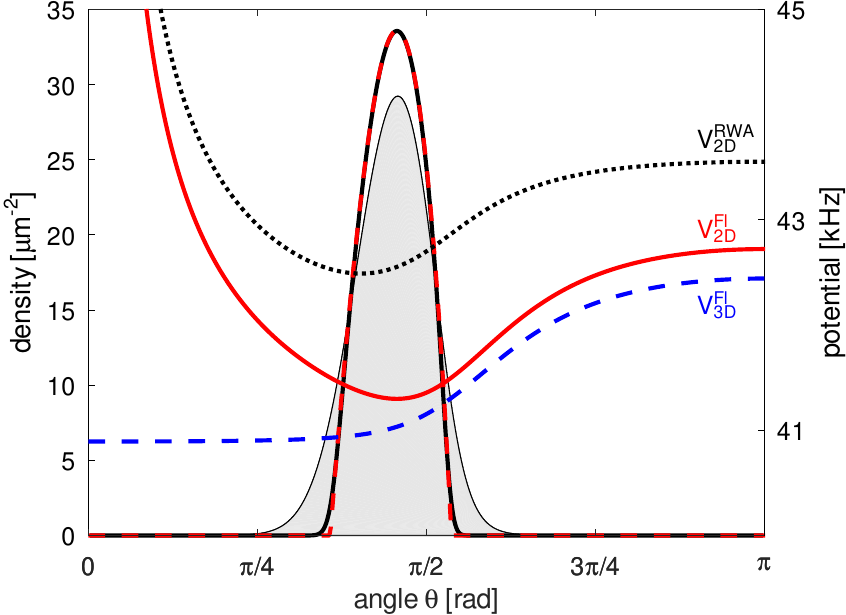}
\caption{\label{fig:sideView}
(color online) 
Right axis: effective potential on the surface $V_{\rm 2D}^{\rm Fl}(\theta)$ (solid red line), $V_{\rm 3D}^{\rm Fl}(r_s(\theta),\theta)$ (dashed blue)
and $V_{\rm 2D}^{\rm RWA}(z)$ (dotted black).
Left axis: surface density computed with the full 3D model (grey shaded area), the 2D semi-classical model (black solid line) and 2D Thomas-Fermi solution (dashed red line).
The trap parameters are: $\omega/(2\pi)=\SI{300}{\kilo\hertz}$, $\Omega_0/(2\pi)=\SI{85.0\pm0.5}{\kilo\hertz}$ and $\alpha/(2\pi)=\SI{7.68\pm0.09}{\kilo\hertz/\micro\metre}$. See text for details.
}
\end{figure}
\Fref{fig:sideView} evidences the difference between the quantum $V_{\rm 2D}^{\rm Fl}(\theta)$ and the classical $V_{\rm 3D}^{\rm Fl}(r_s(\theta),\theta)$ effective potentials, differing by the zero point energy contribution, see equation~\eref{eqn:Vimp}. 
Beyond RWA corrections explain the differences with the simple potential $V_{\rm 2D}^{\rm RWA}(z)$, which nevertheless captures qualitatively the stabilization mechanism. 

In particular, a \emph{classical} particle evolving on the dashed blue potential curve would be pushed towards the top of the spheroid where the rf coupling vanishes, inducing Landau-Zener spin flips. As a consequence, a \emph{classical} particle can not be trapped with this configuration. 
The zero energy contribution provides the necessary barrier preventing the atoms to climb to the top of the ellipsoid.

With the parameters of \fref{fig:sideView}, corresponding to the last column of \fref{fig:ramp}, the local effective trapping frequency along the surface is $\omega_s/(2\pi)\simeq\SI{20}{\hertz}$, and the transverse one varies with $\theta$ over the cloud extent and is equal to $\omega_\perp/(2\pi)=\SI{526}{\hertz}$ at the peak density of the GP groundstate.
The chemical potential of the groundstate is $\mu/h\simeq\SI{450}{\hertz}$ above the potential $V_{\rm 3D}^{\rm Fl}$ at the peak density.
Therefore the quantum gas is well described by an effective two-dimensional model, with a chemical potential $\mu_{\rm 2D}=\mu-\hbar\omega_\perp/2<\hbar\omega_\perp$~\cite{Petrov2000}.
To illustrate this point \fref{fig:sideView} compares the surface density computed with the full 3D simulation (grey shaded area) to effective 2D solutions obtained with the semi-classical potential of equation~\eref{eqn:Vimp}: the 2D GP solution (solid black curve) and 2D Thomas-Fermi profile (dashed red curve), see~\ref{app:2Dmodel}.
Equation~\eref{eqn:Vimp} thus enables a simple and accurate description of the effective 2D dynamics.

\section{Discussion}
When gravity is compensated, any variation of the rf amplitude of technical origin has an important effect.
As a consequence, the rf polarization must be controlled with high accuracy, and the amplitudes and phases of the three antennas are optimized to the $10^{-2}$ level.
After a careful optimization of the rf polarization, the annular gas still presents fluctuations in the density distribution along the annulus of about 30\% rms, with three apparent maxima corresponding to a modulation of the chemical potential of $\sim h\times\SI{300}{\hertz}$ (or $\sim\SI{14}{\nano\kelvin}$).
We have checked that this is due to the small size and distance of the horizontal rf antennas creating the circular polarization, both of order \SI{10}{\milli\metre}, leading to an rf field amplitude modulation at the level of 1\% in a region of size 0.1~mm, with three minima.
We note that for all adiabatic potentials the 2D criterion implies $\mu<\hbar\omega_\perp$, while $\omega_\perp\ll\Omega_0$ to ensure the adiabatic following of the dressed spin state. As $\Omega_0$ enters in the effective potential as a reference energy, see for example~\eref{eqn:Vtoy}, fluctuations of the local chemical potential are of the same order of magnitude than those of $\Omega_0$. For our parameters $\mu/\Omega_0\sim\SI{5e-3}{}$ and targeting a control of the local chemical potential at the 10 percent level would require an unprecedented control of the rf dressing inhomogeneities at the 0.05 percent level.

\begin{figure}[t]
\centering
\includegraphics[width=8.6cm]{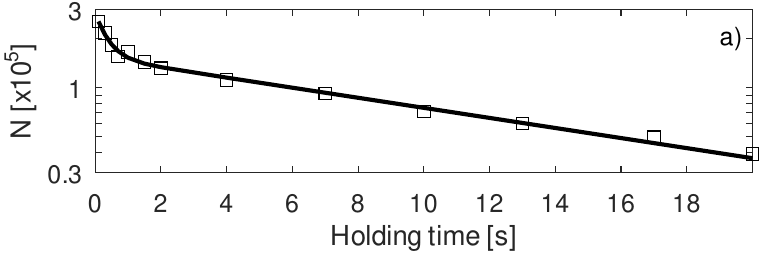}\\
\includegraphics[width=8.6cm]{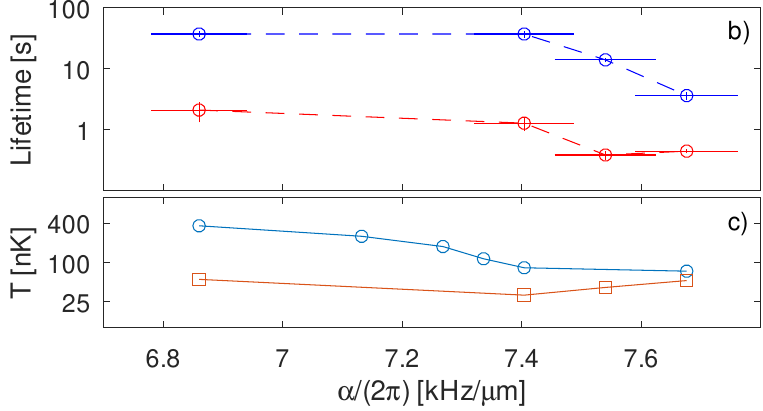}
\caption{\label{fig:lifetime}
(color online) 
a) Atom number as a function of the holding time in the trap, for a gradient of $\alpha/(2\pi)=\SI{7.54\pm0.08}{\kilo\hertz/\micro\metre}$, evidencing a double exponential decay in semilog scale. The black solid line is a fit to the data.
b) Lifetime in the surface trap as a function of the quadrupole trap gradient: the blue/red symbols correspond to the two characteristic timescales, the error bars are the fit uncertainties.
c) Estimated critical temperature (blue circles) for $N=2\times10^5$ atoms and measured upper bound on the temperature (red squares) with the same atom number, see text for details.
}
\end{figure}

For gradients above $\SI{7.6}{\kilo\hertz/\micro\metre}$, the cloud reaches regions where the rf amplitude is too low to ensure efficient rf dressing~\cite{Zobay2001,Zobay2004,Burrows2017}, and we observe increased Landau-Zener losses and a reduced lifetime in the trap, as shown in \fref{fig:lifetime}.
We find that the loss dynamics always follows a double exponential decay as illustrated in \fref{fig:lifetime}a).
We note that the shortest time scale is still much longer than what we expect for three-body losses that remain low for the typical peak density of $\SI{3e13}{\centi\metre^{-3}}$ in the experiment, see~\ref{app:3body}.
Interestingly we have found that the cloud temperature decreases on a timescale compatible with the smallest of the two time scales, and reaches then a stationary value below $\SI{25}{\nano\kelvin}$, see~\ref{app:methods}.
This suggests that an energy-dependent loss mechanism is at play.
Indeed, when the effect of gravity is compensated, thermal atoms can explore a significant fraction of the surface and approach the top of the bubble, where the rf coupling vanishes. 
There they undergo a spin flip to an untrapped state, in a Landau-Zener process, resulting in an energy-dependent filtering of the thermal distribution.
The zero-point energy contribution to the potential acts as a barrier preventing these losses for low energy atoms, allowing to stabilize the gas.

Finally it is worth mentioning that the atoms constrained on the ellipsoidal surface evolve in a highly non separable potential that can not be written as a product of two harmonic oscillators.
This affects the transverse excitation spectrum: the energies do not form a regular ladder, as for the harmonic oscillator.
It would be interesting to study how this effect impacts the quantum gas properties: for example one can expect a modification of the usual equilibrium predictions relying on the harmonic oscillator partition function \cite{Bereta2019,Tononi2019,Moller2020a}.
By diagonalizing the single particule Hamiltonian and computing the density of states $\rho(E)$ we find that, for low energies, it follows a power-law scaling $\rho(E)\propto E^d$, where the exponent decreases smoothly with the gradient from $d=2$ to $d=1/2$ when the ring forms.
Using $\rho(E)$ to compute the ideal Bose gas critical temperature $T_c^0(N)$, we have verified that for the typical atom number in \fref{fig:ramp}, $N=2\times10^5$, the estimated temperature is always below $T_c^0(N)$, see~\ref{app:Tc}, as shown in \fref{fig:lifetime}c).
A more rigorous analysis would require to evaluate the BKT transition temperature as the trap topology changes but goes beyond the scope of this work.

\section{Conclusion}
In conclusion we have reported an attempt to compensate gravity in a shell shaped dressed quadrupole trap, taking advantage of the anti-gravity force induced by the inhomogeneous rf coupling. We have demonstrated that the inhomogeneous transverse confinement plays an important role in determining the equilibrium shape and triggers the appearance of an annular shaped quantum gas.
Therefore we have demonstrated a new method to produce a ring-shaped quantum gas using a particularly simple setup that does not require supplementary oscillating fields~\cite{Sherlock2011} nor optical potentials~\cite{Heathcote2008a}.
We stress that the effect shown here is also relevant in the context of the realization of bubble shaped ensembles in microgravity using adiabatic potentials \cite{Lundblad2019,Tononi2020}.
Our work shows that achieving a homogeneous surface density in these systems seems challenging, as the requirement on magnetic gradient and rf field homogeneity is very high when the relevant energy to be compared to is the chemical potential.
A first step towards improving the homogeneity would be to use larger antennas and/or smaller surfaces.

A consequence of this work is that the quadrupole dressed trap will spontaneously result in an annular trap geometry under microgravity environment.
Once combined with the possibility to tune dynamically the rf polarization it offers an interesting platform to study rotating superfluids in anharmonic traps, on Earth or in space.
Indeed, close to the gravity compensation setting, the harmonic confinement vanishes and the trap at the bottom of the shell has a quartic leading order, leading to new equilibrium vortex distributions in a rotating frame~\cite{Brito2020}.
Furthermore, starting from the annular gas, one can use a small change of the rf polarization to rotate the gas~\cite{Guo2020} and then reduce the gradient to reconnect the cloud, thus implementing a protocol to prepare correlated states~\cite{Roncaglia2011}. 

\ack
We acknowledge enlightening discussions with Maxim Olshanii.
LPL is UMR 7538 of CNRS and Sorbonne Paris Nord University.
We thank USP-COFECUB for support (project Uc Ph 177/19: Out of equilibrium trapped superfluids).
This work has been supported by Région Île-de-France in the framework of DIM SIRTEQ (project Hydrolive).

\appendix
\section{Methods}
\label{app:methods}
\paragraph{Static magnetic field control}
In order to calibrate precisely the gradient of the quadrupole coils, we measure the vertical displacement of the cloud in the dressed quadrupole trap as a function of the dressing frequency $\omega$, from $\SI{300}{\kilo\hertz}$ to $\SI{3}{\mega\hertz}$ and from a linear fit we extract directly the gradient in units of $\SI{}{\kilo\hertz/\micro\metre}$. We load the trap with a reduced coupling $\Omega_0/(2\pi)\simeq\SI{40}{\kilo\hertz}$ such that the atoms are always at the bottom of the ellipsoid, and measure the vertical position of the atoms after a \SI{23}{\milli\second} time-of-flight using an additional imaging axis along an horizontal direction.
An ensemble of large coils along three orthogonal axis allows to cancel the static homogeneous magnetic field at the position of the atoms: therefore the center of mass of the cloud is not displaced in the horizontal plane when the gradient changes.
We repeat this procedure for different gradients in the range $4.14(6)$ to $\SI{8.49\pm0.09}{\kilo\hertz/\micro\metre}$ covering all the data presented in this work and achieve a relative uncertainty of one percent. The experimental values of the gradient given in the main text result from a linear interpolation at any gradient between the measured points.

\begin{figure}[t]
\includegraphics[width=8.6cm]{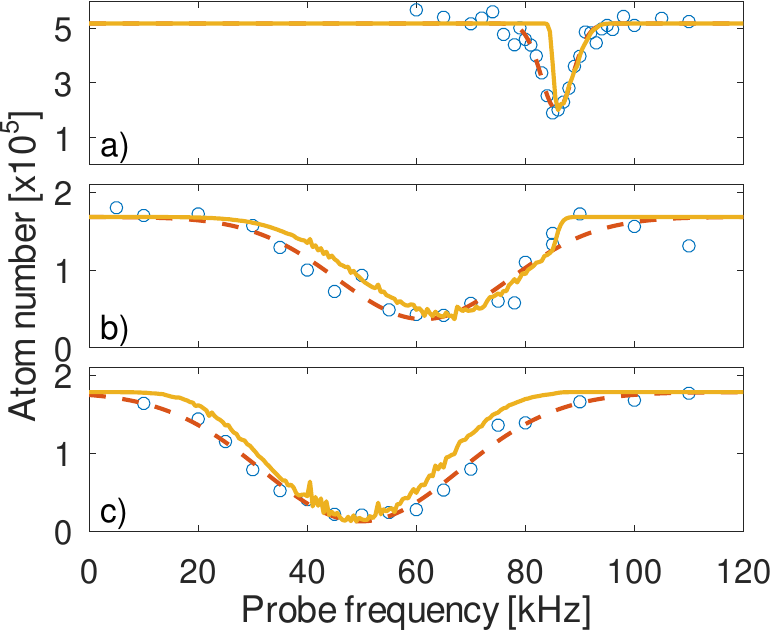}
\caption{\label{fig:spectro}(Color online) rf spectroscopy signal: atom number as a function of the rf probe frequency (open blue circles), Gaussian fit to the data (dashed red curve) and simulation (solid yellow curve). The trap parameters are $\omega=2\pi\times\SI{300}{\kilo\hertz}$, $\Omega_0=2\pi\times\SI{85.0\pm0.5}{\kilo\hertz}$ and $\alpha=2\pi\times\{4.14,7.40,7.54\}~\SI{}{\kilo\hertz/\micro\metre}$ for a), b) and c) respectively.}
\end{figure}

\paragraph{Radio-frequency spectroscopy}
To determine precisely the radio-frequency coupling amplitude we perform radio-frequency spectroscopy~\cite{Hofferberth2007a,Easwaran2010}: using a weak additional rf field, produced by an antenna aligned with the vertical axis, we probe the energy difference between the dressed states at the position of the atoms. When the frequency is resonant this probe field induces losses that are recorded after typically $\SI{500}{\milli\second}$ of weak rf probe pulse. We repeat this measurement for various probe frequencies and record a loss spectrum. At low gradient the resonant frequency is always larger than the effective coupling due to the gravitational sag. A careful comparison with the simulated density distribution is necessary to accurately infer the coupling amplitude. We find $\Omega_0=2\pi\times\SI{85.0\pm0.5}{\kilo\hertz}$, see \fref{fig:spectro}.

For larger gradients, when gravity is overcompensated and the atoms climb on the surface, the local energy difference between dressed states is reduced. As shown in \fref{fig:spectro} this results in a reduction of the resonant frequency and a broadening of the spectrum. The zero-temperature GP simulation captures both effects. Furthermore small variations of the simulation parameters ($\Omega_0$ or $\alpha$) result in noticeable changes of the simulated distribution, allowing us to estimate the uncertainty on the rf coupling amplitude at the level of $\pm\SI{0.5}{\kilo\hertz}$. We note however that several systematic effects can affect this comparison: the experiment is done at small but finite temperature, while the simulation assumes zero temperature, and the spectroscopic signal depends also on the probe polarization, which is not modeled.

\paragraph{Fine tuning of the rf polarization}
The measurements reported in \fref{fig:ramp} and the comparison with numerical simulations assume a perfectly circularly polarized rf field (with respect to the $z$ axis). To achieve this we control the amplitude and the phase of the signals fed to the three dressing antennas. We find that the most sensitive configuration to finely tune the polarization is the over compensated ring trap: any imbalance in the polarization results in density inhomogeneities along the ring. The optimization procedure proceeds as follows: we first roughly equilibrate the amplitudes of the two horizontal plane antennas with a dephasing of $\delta\Phi\sim\pi/2$. This usually results in an inhomogeneous ring with two local density maxima. We then tune the third, vertical axis antenna to balance the atom number between the two maxima by tuning its amplitude $A_z$ and change their relative position on the ring by controlling its phase $\phi_z$, such that we obtain two opposite maxima along one diameter of the ring. Now changing $\delta\Phi$ results in a simultaneous rotation of the two maxima by some angle $\phi_0(\delta\Phi)$. We observe that $\phi_0(\delta\Phi)\sim\arctan{[(\delta\Phi-\delta\Phi_{\rm opt})/\sigma_\Phi]}/2$ where $\delta\Phi_{\rm opt}$ is the optimal phase difference and the width $\sigma_\Phi$ is minimized when the amplitudes of the two horizontal antennas are perfectly balanced. After a few iterations we obtain an almost homogeneous atomic ring, as shown in \fref{fig:ramp}.

\paragraph{In situ imaging}
We use a homemade four lenses imaging objective attached to the camera. The depth of view is about $\SI{100}{\micro\metre}$, larger than the vertical distance traveled by the atoms in the picture series of \fref{fig:ramp}. The resolution is $\SI{4}{\micro\metre}$ (Rayleigh criterion), limited by the numerical aperture $\sim 0.1$. After alignment and focus adjustment using a triaxial translation stage, we take several pictures of a small cloud at the bottom of the trap while moving the imaging system along an horizontal axis. From a fit of the center of mass of the cloud position as a function of the displacement, measured on the translation stage, we obtain a magnification of $7.78$.

\paragraph{Temperature estimation}
To estimate the atomic cloud temperature, one could use the in situ density profile and the knowledge of the equation of state~\cite{Yefsah2011,Hung2011a}. Unfortunately we cannot rely on this method as the in situ pictures are significantly affected by the limited resolution of our low numerical aperture ($\sim0.1$) objective. Therefore we use a time-of-flight measurement, that allows to estimate the velocity distribution.
However this method is not well adapted to our experiment: as the initial cloud is very far from a simple harmonically trapped ensemble, we cannot rely on a simple model to describe the cloud expansion. It has been shown that an expansion from ring or bubble shaped traps results in subtle interference phenomena. Furthermore we are limited in our vacuum cell to time-of-flight $t_{\rm tof}\leq\SI{30}{\milli\second}$. Even if we assume a ballistic expansion of the cloud, purely driven by the initial thermal velocity distribution, such that the rms size $\sigma$ obeys $\sigma^2=\sigma_0^2+k_BT t_{\rm tof}^2/M$, the temperature is accurately determined when the second term is larger than the first one. This results in a limit temperature sensitivity $T\geq\sigma_0^2M/(k_Bt_{\rm tof}^2)\simeq\SI{20}{\nano\kelvin}$, where we assumed $\sigma_0^2\simeq r_0^2$ for a ring shaped distribution.

We note also that for a low-dimensional system the expansion along the initially frozen degree of freedom results typically in a similar ballistic expansion, governed by the initial velocity fluctuations $v^2\sim\hbar\omega_\perp/M$. This also sets a limit on the temperature measurement: $T\geq \hbar\omega_\perp/k_B\simeq\SI{24}{\nano\kelvin}$ for our parameters.

Despite the complex expansion dynamics, we observe that the density profile after time-of-flight displays a bi-modal shape, with a background Gaussian pedestal. Using a simple Gaussian fit we extract the rms size and deduce an upper bound for the temperature $T_{\rm max}$. For all the data presented in this work we find $T_{\rm max}$ varying from \SI{60}{\nano\kelvin}, at low gradients, to \SI{25}{\nano\kelvin}, when the ring shape appears, probably limited by the above mentioned factors for the lowest temperatures.

\section{Derivation of the effective two-dimensional model}%Dimensional reduction}
\label{app:2Dmodel}
We provide here a short summary of dimensional reduction on a surface~\cite{DaCosta1981,Kaplan1997,Sandin2017a,Schwartz2006a}, adapted to the geometry of the experiment reported in the main text. The full mathematical derivation will be discussed elsewhere and we focus only on the key ingredients. In particular we have verified that the contribution of the surface curvature itself is small compared to the inhomogeneous transverse confinement and does not play a key role. Therefore we start by recalling that the mean field groundstate is found by solving the three-dimensional Gross-Pitaevskii (GP) equation:
\[
\mu\psi=\left(-\frac{\hbar^2}{2M}\Delta+V(\bm{r})+g|\psi|^2\right)\psi,
\]
where the wavefunction $\psi\equiv\psi(\bm{r})$ is normalized to the number of particles: $N=\int d^3\bm{r}\,|\psi|^2$, $\mu$ is the chemical potential, $M$ is the atomic mass, and $g=4\pi a_s\hbar^2/M$ is the two-body interaction strength, with $a_s$ the low energy $s$-wave scattering length.

When one dimension is strongly confined by a tight harmonic oscillator of frequency $\omega_\perp$, such that the atoms occupy only the groundstate along this dimension, an effective two-dimensional GP equation can be derived~\cite{Petrov2000}:
\[
\mu\psi_s=\left(-\frac{\hbar^2}{2M}\Delta_{s}+\frac{\hbar\omega_\perp}{2}+V_{s}+\frac{g}{\sqrt{2\pi}\sigma}|\psi_s|^2\right)\psi_s,
\]
where $\psi_s$, $\Delta_s$, $V_s$ are the wavefunction, Laplacian and potential, restricted onto the surface, respectively, and $\sigma=\sqrt{\hbar/(M\omega_\perp)}$ is the length scale associated to the transverse confinement $\omega_\perp$.

Since we neglect here all curvature effects, we may connect directly this equation with the notations of the main text:
\begin{eqnarray*}
\mu\psi_s&=&-\frac{\hbar^2}{2M}\frac{1}{r_s(\theta)^2\sin{\theta}}\frac{\partial}{\partial\theta}\left(\sin{\theta}\frac{\partial \psi_s}{\partial\theta}\right)\\
&&+\left(V_{\rm 2D}^{\rm Fl}(\theta)+\frac{g}{\sqrt{2\pi}\sigma(\theta)}|\psi_s|^2\right)\psi_s,
\end{eqnarray*}
where $V_{\rm 2D}^{\rm Fl}(\theta)\equiv V(r_s(\theta),\theta)+\hbar\omega_\perp(\theta)/2$.
We then obtain the Thomas-Fermi solution by neglecting the kinetic energy term resulting in:
\[
|\psi_s|^2=\frac{\sqrt{2\pi}\sigma(\theta)}{g}\left(\mu-V_{\rm 2D}^{\rm Fl}(\theta)\right).
\]
As shown in the main text this simple form captures the main features of the dimensional reduction.

\section{Estimation of three-body losses}
\label{app:3body}
In order to estimate the three-body losses, one has to solve the equation $\dot{N}=-K_3\int \bm{dr}\,n(\bm{r},t)^2$, where the density profile $n(\bm{r},t)$ must be computed self-consistently and $K_3=\SI{6e-42}{\metre^{6}}$ for $^{87}$Rb.
On the one hand, if we assume a three-dimensional ring geometry, with a Thomas-Fermi density profile $n_{3D}(r,z)=\frac{\mu_{3D}}{g}\left(1-\frac{(r-r_0)^2}{R^2}-\frac{z^2}{R_z^2}\right)$, where $R=\sqrt{2\mu/(M\omega_r^2)}$, $R_z=\sqrt{2\mu/(M\omega_z^2)}$ are the horizontal and vertical Thomas-Fermi radii respectively, and the chemical potential is $\mu_{3D}=\hbar\sqrt{\omega_r\omega_z}\sqrt{2Na_s/(\pi r_0)}$~\cite{Morizot2006}, we find that three-body losses obey: $N(t)=N_0/(1+\gamma_{3D}N_0 t)$, where:
\[
\gamma_{3D}=\frac{K_3}{16\pi^3a_sr_0a_r^2a_z^2},
\]
where $a_{r,z}=\sqrt{\hbar/(M\omega_{r,z})}$ are the harmonic oscillator length scales.
On the other hand, if we assume a two-dimensional ring geometry, with density profile $n_{2D}(r,z)=\frac{\mu}{g_{2D}}\left(1-\frac{z^2}{R_z^2}\right)\frac{e^{-\frac{(r-r_0)^2}{a_r^2}}}{\sqrt{\pi}a_r}$,
where $g_{2D}=g/(\sqrt{2\pi}a_r)$, the chemical potential is $\mu_{2D}=\hbar(\omega_r\omega_z^2)^{1/3}(3Na_s/r_0)^{2/3}/(2(2\pi)^{1/3})$, and we find that three-body losses obey: $N(t)=N_0/(1+\gamma_{2D}N_0^{4/3}t)^{3/4}$, where:
\[
\gamma_{2D}=K_3\frac{3^{5/6}(a_sr_0^2a_r^2a_z^4)^{-2/3}}{35\times2^{2/3}\pi^{8/3}}.
\]
Remarkably we find that the value of $\gamma_{2D}$ is not changed if one assumes that the dimensional reduction occurs along the radial coordinate (instead of the vertical one).

\begin{figure}[t]
\includegraphics[width=8.6cm]{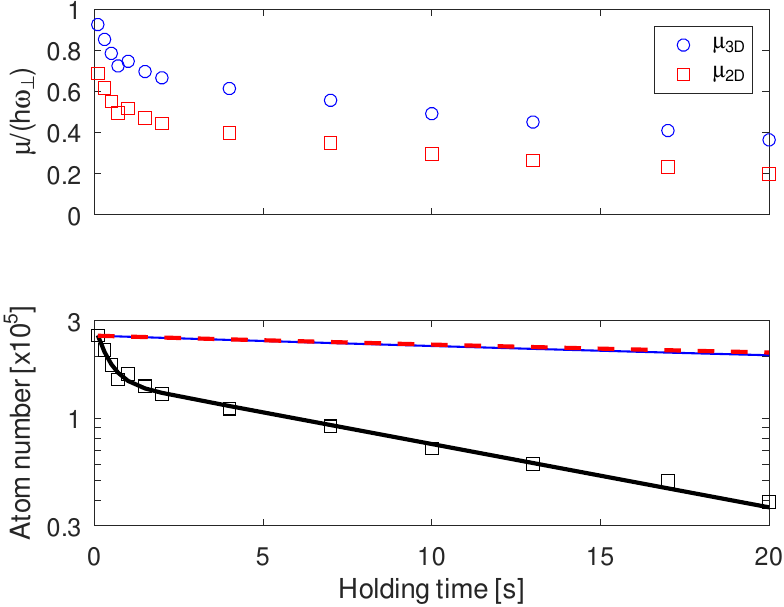}
\caption{\label{fig:3body}
(Color online) Analysis of the dimensionality and loss rate for a gradient of $\alpha/(2\pi)=\SI{7.54(8)}{\kilo\hertz/\micro\metre}$. Top graph: ratio of the chemical potential to the strong confinement energy $\mu/\hbar\omega_\perp$ computed using a 3D (2D) model, blue circles (red squares). Bottom graph: measured loss dynamics (square symbols) and double exponential fit (solid black curve), compared to expected three-body losses using a 3D (2D) model, matching the measured atom number, solid blue (dashed red) curves. See text for details.
}
\end{figure}

\Fref{fig:3body} shows the result for the three-body loss rate estimation for the shell potential with over-compensated gravity, using a very simple model of harmonic ring trap, with frequencies $\omega_\perp/(2\pi)=\omega_r/(2\pi)\sim\SI{500}{\hertz}$ and $\omega_z/(2\pi)\sim\SI{20}{\hertz}$, as discussed above, and the measured atom number as an input. It confirms that the gas is in the two-dimensional regime $\mu<\hbar\omega_\perp$ and that the loss dynamics cannot be explained by three-body losses. We note that the harmonic ring trap model is a very crude approximation of the effective 2D potential on the shell (due to a significant anharmonicity), leading to an overestimation of the peak density and hence of the three-body loss rate.

\section{Analytic formulas with rotating-wave approximation}
As mentioned in the main text the RWA formulas are useful to derive analytic formulas. In particular, using equation~\eref{eqn:potential}, that we recall here:
\[
V_{\rm 3D}^{\rm RWA}(\rho,\phi,z)=\hbar\sqrt{(\alpha\ell-\omega)^2+\Omega(\rho,\phi,z)^2}+Mgz,
\]
one can evaluate with good accuracy the transverse confinement frequency.
Assuming a strong transverse confinement, such that the correction due to the gravitational potential is small, the atoms are localized close to the resonant surface $\ell=r_0$. Locally, the tangent plane to this surface is given by the angle $\beta$ such that $\tan{\beta}=-\rho/(4z)=-\tan{[\theta]}/4$ and the transverse confinement is given by:
\[
\omega_\perp(z)^2=\frac{1}{M}\left.\frac{d^2}{du^2}V_{\rm 3D}^{\rm RWA}(\rho-u\sin{\beta},z+u\cos{\beta},\phi)\right|_{u=0},
\]
where $\rho=\sqrt{r_0^2-4z^2}$ on the surface. A lengthy but straightforward calculation gives:
\begin{equation}
\omega_\perp(z)\simeq\alpha(z)\sqrt{\frac{\hbar}{M\Omega(z)}},
\label{eqn:omega_approx}
\end{equation}
where $\alpha(z)=\alpha\sqrt{1+12z^2/r_0^2}$ is the gradient along the normal to the surface, $\Omega(z)=\Omega_0/2\times(1-2z/r_0)$ is the rf coupling on the surface and we have neglected a (small) correction in \eref{eqn:omega_approx} of order $(\Omega_0/\omega)^2$.

\begin{figure}[t]
\includegraphics[width=8.6cm]{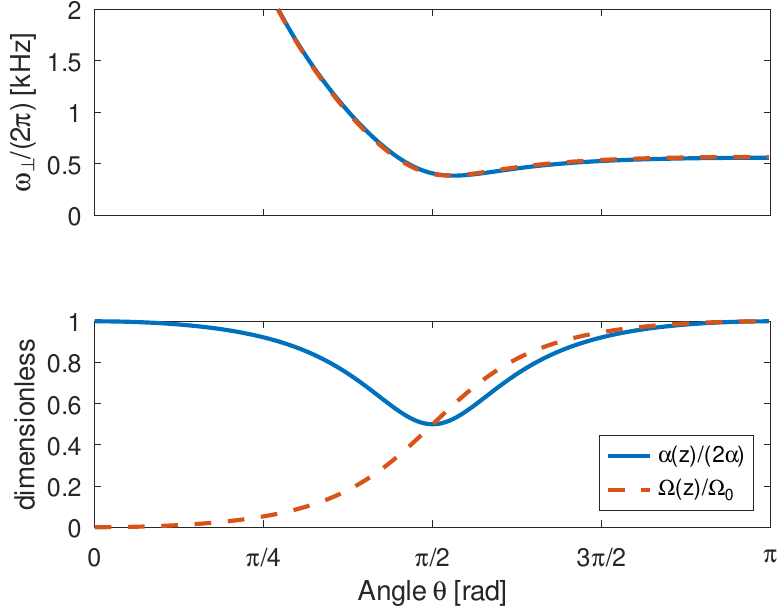}
\caption{\label{fig:approxRWA}
(Color online) Top graph: transverse local confinement frequency as a function of $\theta$, computed with the full Floquet potential (solid blue line) or with the analytical RWA result of equation~\eref{eqn:omega_approx} (dashed red line). Bottom graph: local transverse gradient $\alpha(z)$, normalized to the maximum gradient $2\alpha$ (solid blue curve) and local rf coupling $\Omega(z)$, normalized to the maximum coupling $\Omega_0$ (red dashed curve), computed on the resonant surface $z=r_0\cos{\theta}/\sqrt{1+3\cos^2\theta}$.
}
\end{figure}

\Fref{fig:approxRWA} compares the prediction of equation~\eref{eqn:omega_approx} to an exact numerical computation, as explained in the main text. The agreement is remarkably good, at the level of a few percents for the range displayed. The second panel shows the variations of the gradient and rf coupling on the resonant surface. These simple analytical formulas are useful to understand our results, however we note that the accuracy needed to match quantitatively the experimental results is obtained only with more involved numerical methods.

\section{Critical temperature}
\label{app:Tc}
To estimate the critical temperature for Bose-Einstein condensation in our trap, for different gradients, we compute the ideal gas result $T_c^0$. As the trap geometry is highly non trivial we compute it using the exact single particle spectrum, obtained by a numerical diagonalization of the Hamiltonian. We make use of the cylindrical symmetry about the vertical axis to simplify the spectrum computation, introducing the angular momentum $m$ quantum number.
Then the number of atoms in the excited states is:
\[
N^\prime=\sum_{(m,n)\neq(0,0)}%\substack{m\geq0,n\\ (m,n)\neq (0,0)}}
\frac{g_m}{\exp{\left[\frac{\epsilon_{m,n}-\mu}{k_BT}\right]}-1},
\]
where $g_0=1$ and $g_m=2$ for $m>0$ is the degeneracy of the state with energy $\epsilon_{m,n}$ and $n$ labels the single particle eigenstates. In this expression the sum runs over all the states, except the groundstate $\epsilon_{0,0}$. In order to have a reasonable computation time, we include states up to energies $E_{\rm max}=\epsilon_{0,0}+k_B\times\SI{144}{\nano\kelvin}$. We then evaluate the sum when $\mu\to\epsilon_{0,0}$ to obtain the critical atom number as a function of the temperature or equivalently $T_c^0(N)$.

\begin{figure}[t]
\includegraphics[width=8.6cm]{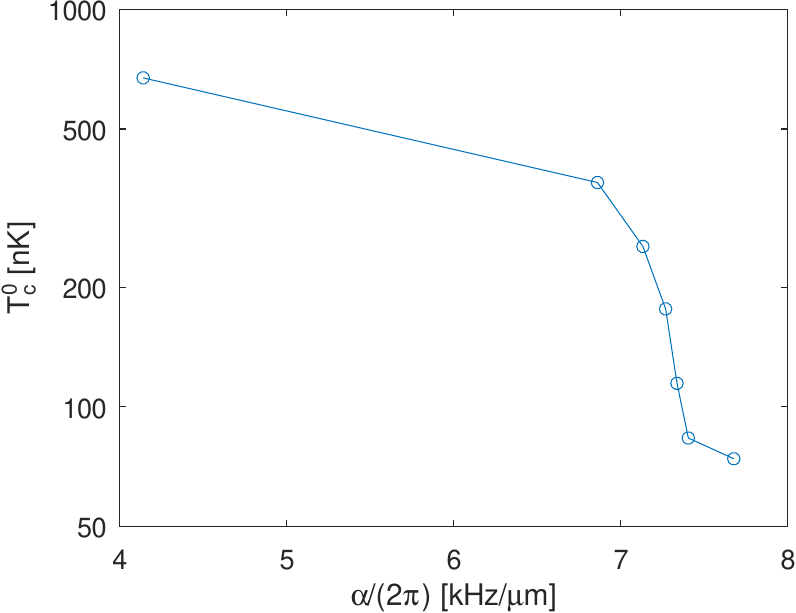}
\caption{\label{fig:EOS}
(Color online) Critical temperature of the ideal Bose gas for $N=10^5$ atoms, as a function of the gradient and for rf dressing with $\omega/(2\pi)=\SI{300}{\kilo\hertz}$ and $\Omega_0/(2\pi)=\SI{85.0\pm0.5}{\kilo\hertz}$.
}
\end{figure}

From the knowledge of the single particle spectrum we can also evaluate the density of states $\rho(E)$ and study how it varies when we increase the gradient. We find that $\rho(E)\propto E^2$ at low gradients, as expected for a 3D harmonic oscillator, $\rho(E)\propto E^{1}$ at moderate gradient, characteristic of a 2D harmonic oscillator and then evolves towards $\rho(E)\propto\sqrt{E}$ as the gradient increases, evidencing the change of topology. As the low energy density of states increases with the gradient, the critical temperature decreases but remains above the estimated upper bound for the temperature in the experiment. \Fref{fig:EOS} shows the dependence of $T_c^0(N)$ with the gradient, for a total atom number of $N=10^5$.

\section{Floquet expansion}
\label{app:Floquet}
We briefly summarize here how we perform the Floquet expansion adapted to our study. We start by recalling the Hamiltonian for an atom of total spin $\hat{\bm{F}}$ in a combination of a static inhomogeneous magnetic field and oscillating homogeneous magnetic field:
\[
\hat{H}_0=\frac{\hat{\bm{p}}^2}{2M}+\left[\omega_0(\bm{r})+\Omega_z(\bm{r},t)\right]\hat{F}_z+\Omega_+(\bm{r},t)\hat{F}_++\Omega_-(\bm{r},t)\hat{F}_-,
\]
where $\omega_0(\bm{r})$ is the local Larmor frequency, due to the static field and $\Omega_{\pm,z}(\bm{r},t)$ are the couplings induced by the oscillating field, in the $\pi$, $\sigma^\pm$ polarizations and $\hat{F}_\pm=\hat{F}_x\pm i\hat{F}_y$ are the raising and lowering operators. Here we assume that the atomic spin adiabatically follows the inhomogeneous static magnetic field and that the rf-dressing coils are large enough such that the rf-field can be considered as homogeneous. However, even if the rf field is homogeneous, the coupling $\Omega_{z,\pm}(\bm{r},t)$ are not because their relative orientation with respect to the inhomogeneous static field depend explicitly on the position. To transform $\hat{H}_0$ in an explicitly time-independent Hamiltonian we proceed in two steps: first we treat the $\pi$ polarization term exactly following the approach of Ref.~\cite{Haroche1970} resulting in a renormalization of the $\sigma^\pm$ couplings by Bessel functions weights and second we perform the Floquet expansion, looking for a solution of the form~\cite{Shirley1965,Hofferberth2007a}:
\[
\ket{\psi}=\sum_n e^{i(n\omega-E/\hbar)t}\ket{\psi_n},
\]
resulting in an infinite system of coupled equations:
\[
E\ket{\psi_n}=\hat{D}_n(\bm{r})\ket{\psi_n}+\sum_{k\neq0}\hat{V}_k(\bm{r})\ket{\psi_{k+n}},
\]
where $\hat{D}_n(\bm{r})=\hat{\bm{p}}^2/(2M)+n\hbar\omega\hat{I}-\delta(\bm{r})\hat{F}_z+\hat{V}_0(\bm{r})$, $\delta=\omega-\omega_0(\bm{r})$ being the local detuning and the coupling terms are
\[
\hat{V}_k(\bm{r})=\sum_l\left[\tilde{\Omega}^{(l+1+k)}_+(\bm{r})c_l(\bm{r})\hat{F}_++\tilde{\Omega}^{(l+1-k)}_-(\bm{r})c_l(\bm{r})^*\hat{F}_-\right],
\]
where $\tilde{\Omega}_\pm^{(l)}(\bm{r})$ is the $l$-th harmonic of the rf coupling $\Omega_\pm(\bm{r},t)$ and $c_n(\bm{r})=J_n[\Omega_0\rho/\omega\ell]e^{-in\phi}$ is the Bessel weight due to the $\pi$ polarized rf component. To obtain finally the effective dressing potential $V_{\rm 3D}^{\rm Fl}(\bm{r})$ we truncate the Floquet expansion to $|n|\leq2$, find the eigenvalues by a standard diagonalization algorithm and repeat this procedure for each needed spatial position. We have verified that using a larger Floquet Hamiltonian does not change the results, meaning that the eigenvalues are correctly evaluated with the second order expansion.

\section*{References}
\bibliographystyle{iopart-num}
%\bibliography{biblio}

\providecommand{\newblock}{}
\begin{thebibliography}{}
\expandafter\ifx\csname url\endcsname\relax
  \def\url#1{{\tt #1}}\fi
\expandafter\ifx\csname urlprefix\endcsname\relax\def\urlprefix{URL }\fi
\providecommand{\eprint}[2][]{\url{#2}}

\end{thebibliography}


\begin{thebibliography}{10}
\expandafter\ifx\csname url\endcsname\relax
  \def\url#1{{\tt #1}}\fi
\expandafter\ifx\csname urlprefix\endcsname\relax\def\urlprefix{URL }\fi
\providecommand{\eprint}[2][]{\url{#2}}

\bibitem{DaCosta1981}
Da~Costa R~C~T 1981 {\em Phys. Rev. A\/} {\bf 23} 1982--1987

\bibitem{Krinner2015}
Krinner S, Stadler D, Husmann D, Brantut J~P and Esslinger T 2015 {\em
  Nature\/} {\bf 517} 64--67

\bibitem{Imry1998}
Imry Y 1998 {\em Physica Scripta\/} {\bf T76} 171 ISSN 0031-8949
  (\textit{Preprint} \eprint{9807306})
  \urlprefix\url{https://iopscience.iop.org/article/10.1238/Physica.Topical.076a00171}

\bibitem{Hall1879}
Hall E~H 1879 {\em American Journal of Mathematics\/} {\bf 2} 287 ISSN 00029327
  \urlprefix\url{https://www.jstor.org/stable/2369245?origin=crossref}

\bibitem{Kaplan1997}
Kaplan L, Maitra N~T and Heller E~J 1997 {\em Phys. Rev. A\/} {\bf 56}
  2592--2599

\bibitem{Sandin2017a}
Sandin P, {\"{O}}gren M, Gulliksson M, Smyrnakis J, Magiropoulos M and
  Kavoulakis G~M 2017 {\em Phys. Rev. E\/} {\bf 95} 012142

\bibitem{Schwartz2006a}
Schwartz S, Cozzini M, Menotti C, Carusotto I, Bouyer P and Stringari S 2006
  {\em New J. Phys.\/} {\bf 8} 162--162

\bibitem{Gorlitz2001}
G{\"{o}}rlitz A, Vogels J~M, Leanhardt A~E, Raman C, Gustavson T~L, Abo-Shaeer
  J~R, Chikkatur A~P, Gupta S, Inouye S, Rosenband T and Ketterle W 2001 {\em
  Phys. Rev. Lett.\/} {\bf 87} 130402

\bibitem{QGLD2003}
Pricoupenko L, Perrin H and Olshanii M (eds) 2004 {\em Quantum Gases in Low
  Dimensions\/} vol 116 (J. Phys. IV France)
  \urlprefix\url{https://jp4.journaldephysique.org/articles/jp4/abs/2004/04/contents/contents.html}

\bibitem{Kinoshita2006b}
Kinoshita T, Wenger T and Weiss D~S 2006 {\em Nature\/} {\bf 440} 900--903

\bibitem{Hadzibabic2006}
Hadzibabic Z, Kr{\"{u}}ger P, Cheneau M, Battelier B and Dalibard J 2006 {\em
  Nature\/} {\bf 441} 1118--1121

\bibitem{Fletcher2015}
Fletcher R~J, Robert~de Saint-Vincent M, Man J, Navon N, Smith R~P, Viebahn
  K~G~H and Hadzibabic Z 2015 {\em Phys. Rev. Lett.\/} {\bf 114} 255302

\bibitem{Mancini2015}
Mancini M, Pagano G, Cappellini G, Livi L, Rider M, Catani J, Sias C, Zoller P,
  Inguscio M, Dalmonte M and Fallani L 2015 {\em Science\/} {\bf 349}
  1510--1513

\bibitem{Chalopin2020}
Chalopin T, Satoor T, Evrard A, Makhalov V, Dalibard J, Lopes R and Nascimbene
  S 2020 {\em Nat. Phys.\/} {\bf 16} 1017--1021

\bibitem{Saint-Jalm2019a}
Saint-Jalm R, Castilho P~C, {Le Cerf}, Bakkali-Hassani B, Ville J~L, Nascimbene
  S, Beugnon J and Dalibard J 2019 {\em Phys. Rev. X\/} {\bf 9} 21035

\bibitem{Shi2020}
Shi Z~Y, Gao C and Zhai H 2020 Idealized hydrodynamics (\textit{Preprint}
  \eprint{2011.01415})

\bibitem{Olshanii2021}
Olshanii M, Deshommes D, Torrents J, Gonchenko M, Dunjko V and Astrakharchik
  G~E 2021 Triangular {Gross-Pitaevskii} breathers and {Damski-Chandrasekhar}
  shock waves (\textit{Preprint} \eprint{2102.12184})

\bibitem{Olshanii1998}
Olshanii M 1998 {\em Phys. Rev. Lett.\/} {\bf 81} 938--941

\bibitem{Olshanii2010}
Olshanii M, Perrin H and Lorent V 2010 {\em Phys. Rev. Lett.\/} {\bf 105}
  095302

\bibitem{Hung2011a}
Hung C~L, Zhang X, Gemelke N and Chin C 2011 {\em Nature\/} {\bf 470} 236--239

\bibitem{Merloti2013a}
Merloti K, Dubessy R, Longchambon L, Perrin A, Pottie P~E~P~E, Lorent V and
  Perrin H 2013 {\em New J. Phys.\/} {\bf 15} 033007

\bibitem{Bereta2019}
Bereta S~J, Madeira L, Bagnato V~S and Caracanhas M~A 2019 {\em Am. J. Phys.\/}
  {\bf 87} 924--934

\bibitem{Tononi2019}
Tononi A and Salasnich L 2019 {\em Phys. Rev. Lett.\/} {\bf 123} 160403

\bibitem{Tononi2020}
Tononi A, Cinti F and Salasnich L 2020 {\em Phys. Rev. Lett.\/} {\bf 125}
  010402

\bibitem{Moller2020a}
M{\'{o}}ller N~S, dos Santos F~E~A, Bagnato V~S and Pelster A 2020 {\em New
  Journal of Physics\/} {\bf 22} 063059
  \urlprefix\url{https://iopscience.iop.org/article/10.1088/1367-2630/ab91fb}

\bibitem{Bereta2021}
Bereta S~J, Caracanhas M~A and Fetter A~L 2021 {\em Phys. Rev. A\/} {\bf
  103}(5) 053306
  \urlprefix\url{https://link.aps.org/doi/10.1103/PhysRevA.103.053306}

\bibitem{Lundblad2019}
Lundblad N, Carollo R~A, Lannert C, Gold M~J, Jiang X, Paseltiner D, Sergay N
  and Aveline D~C 2019 {\em npj Microgravity\/} {\bf 5} 30

\bibitem{Garraway2016a}
Garraway B~M and Perrin H 2016 {\em J. Phys. B\/} {\bf 49} 172001

\bibitem{Schumm2005b}
Schumm T, Hofferberth S, Andersson L~M, Wildermuth S, Groth S, Bar-Joseph I,
  Schmiedmayer J and Kr{\"u}ger P 2005 {\em Nat. Phys.\/} {\bf 1} 57

\bibitem{Hofferberth2006}
Hofferberth S, Lesanovsky I, Fischer B, Verdu J and Schmiedmayer J 2006 {\em
  Nat. Phys.\/} {\bf 2} 710--716

\bibitem{Lesanovsky2006}
Lesanovsky I, Schumm T, Hofferberth S, Andersson L~M, Kr{\"{u}}ger P and
  Schmiedmayer J 2006 {\em Phys. Rev. A\/} {\bf 73} 033619

\bibitem{Barker2020}
Barker A~J, Sunami S, Garrick D, Beregi A, Luksch K, Bentine E and Foot C~J
  2020 {\em New J. Phys.\/} {\bf 22} 103040

\bibitem{Colombe2004a}
Colombe Y, Knyazchyan E, Morizot O, Mercier B, Lorent V and Perrin H 2004 {\em
  Eur. Phys. Lett.\/} {\bf 67} 593--599

\bibitem{Merloti2013b}
Merloti K, Dubessy R, Longchambon L, Olshanii M and Perrin H 2013 {\em Phys.
  Rev. A\/} {\bf 88} 061603

\bibitem{Dubessy2014}
Dubessy R, {De Rossi} C, Badr T, Longchambon L and Perrin H 2014 {\em New J.
  Phys.\/} {\bf 16} 122001

\bibitem{DeRossi2016}
{De Rossi} C, Dubessy R, Merloti K, De~Go{\"{e}}r~de Herve M, Badr T, Perrin A,
  Longchambon L and Perrin H 2016 {\em New J. Phys.\/} {\bf 18} 062001

\bibitem{Guo2020}
Guo Y, Dubessy R, De~Go{\"{e}}r~de Herve M, Kumar A, Badr T, Perrin A,
  Longchambon L and Perrin H 2020 {\em Phys. Rev. Lett.\/} {\bf 124} 025301

\bibitem{Lesanovsky2007a}
Lesanovsky I and {von Klitzing} W 2007 {\em Phys. Rev. Lett.\/} {\bf 99} 083001

\bibitem{Sherlock2011}
Sherlock B~E, Gildemeister M, Owen E, Nugent E and Foot C~J 2011 {\em Phys.
  Rev. A\/} {\bf 83} 043408

\bibitem{Harte2018}
Harte T~L, Bentine E, Luksch K, Barker A~J, Trypogeorgos D, Yuen B and Foot C~J
  2018 {\em Phys. Rev. A\/} {\bf 97} 013616

\bibitem{Navez2016}
Navez P, Pandey S, Mas H, Poulios K, Fernholz T and {von Klitzing} W 2016 {\em
  New J. Phys.\/} {\bf 18} 075014

\bibitem{Pandey2019}
Pandey S, Mas H, Drougakis G, Thekkeppatt P, Bolpasi V, Vasilakis G, Poulios K
  and {von Klitzing} W 2019 {\em Nature\/} {\bf 570} 205--209

\bibitem{Dubessy2012a}
Dubessy R, Merloti K, Longchambon L, Pottie P~E, Liennard T, Perrin A, Lorent V
  and Perrin H 2012 {\em Phys. Rev. A\/} {\bf 85} 013643

\bibitem{Zobay2001}
Zobay O and Garraway B~M 2001 {\em Phys. Rev. Lett.\/} {\bf 86} 1195--1198

\bibitem{Perrin2017}
Perrin H and Garraway B~M 2017 Chapter four - {T}rapping atoms with radio
  frequency adiabatic potentials ({\em Advances In Atomic, Molecular, and
  Optical Physics\/} vol~66) ed Arimondo E, Lin C~C and Yelin S~F (Academic
  Press) pp 181--262

\bibitem{Shirley1965}
Shirley J~H 1965 {\em Phys. Rev.\/} {\bf 138}(4B) B979--B987
  \urlprefix\url{https://link.aps.org/doi/10.1103/PhysRev.138.B979}

\bibitem{Hofferberth2007a}
Hofferberth S, Fischer B, Schumm T, Schmiedmayer J and Lesanovsky I 2007 {\em
  Phys. Rev. A\/} {\bf 76} 013401

\bibitem{Antoine2017}
Antoine X, Besse C, Duboscq R and Rispoli V 2017 {\em Comput. Phys. Commun.\/}
  {\bf 219} 70--78

\bibitem{Petrov2000}
Petrov D~S, Holzmann M and Shlyapnikov G~V 2000 {\em Phys. Rev. Lett.\/} {\bf
  84} 2551--2555 ISSN 0031-9007
  \urlprefix\url{http://link.aps.org/doi/10.1103/PhysRevB.54.7593
  http://link.aps.org/doi/10.1103/PhysRevLett.84.2551}

\bibitem{Zobay2004}
Zobay O and Garraway B~M 2004 {\em Phys. Rev. A\/} {\bf 69} 023605

\bibitem{Burrows2017}
Burrows K~A, Perrin H and Garraway B~M 2017 {\em Phys. Rev. A\/} {\bf 96}(2)
  023429 \urlprefix\url{https://link.aps.org/doi/10.1103/PhysRevA.96.023429}

\bibitem{Heathcote2008a}
Heathcote W~H, Nugent E, Sheard B~T and Foot C~J 2008 {\em New J. Phys.\/} {\bf
  10} 043012

\bibitem{Brito2020}
Brito L, Andriati A, Tomio L and Gammal A 2020 {\em Phys. Rev. A\/} {\bf
  102}(6) 063330

\bibitem{Roncaglia2011}
Roncaglia M, Rizzi M and Dalibard J 2011 {\em Sci. Rep.\/} {\bf 1} 1--7

\bibitem{Easwaran2010}
Easwaran R~K, Longchambon L, Pottie P~E, Lorent V, Perrin H and Garraway B~M
  2010 {\em J. Phys. B\/} {\bf 43} 065302

\bibitem{Yefsah2011}
Yefsah T, Desbuquois R, Chomaz L, G{\"u}nter K~J and Dalibard J 2011 {\em Phys.
  Rev. Lett.\/} {\bf 107}(13) 130401
  \urlprefix\url{http://link.aps.org/doi/10.1103/PhysRevLett.107.130401}

\bibitem{Morizot2006}
Morizot O, Colombe Y, Lorent V, Perrin H and Garraway B~M 2006 {\em Phys. Rev.
  A\/} {\bf 74} 023617

\bibitem{Haroche1970}
Haroche S, Cohen-Tannoudji C, Audoin C and Schermann J~P 1970 {\em Phys. Rev.
  Lett.\/} {\bf 24} 861--864

\end{thebibliography}
\providecommand{\newblock}{}

\end{document}